\documentclass{article}

\usepackage[utf8]{inputenc}
\usepackage{graphicx}
\usepackage{caption}
\usepackage{subcaption}
\usepackage{bm}
\usepackage{amsmath}
\usepackage{authblk}

\title{Identification of the nature of dynamical systems with recurrence plots and convolution neural networks: A preliminary test}

\author[1]{Daniel Han}
\author[2]{Giuseppe Orlando\footnote{giuseppe.orlando@uniba.i}} 
\author[1]{Sergei Fedotov\footnote{sergei.fedotov@manchester.ac.uk}}

\affil[1]{Department of Mathematics, University of Manchester, M13 9PY, United Kingdom}
\affil[2]{Universit\`{a} degli Studi di Bari ``Aldo Moro'',
	Department of~Economics and Finance,
	Largo Abbazia S.~Scolastica,
	Bari, I-70124 Italy}

\date{\today}

\begin{document}
	
	\maketitle
	
	\begin{abstract}
		In this study, we present  a method for classifying dynamical systems using a hybrid approach involving recurrence plots and a convolution neural network (CNN). This is performed by obtaining the recurrence matrix of a time series generated from a given dynamical system and then using a CNN to classify the related dynamics observed from the recurrence matrix. We consider three broad classes of dynamics: chaotic, periodic, and stochastic. Using a relatively simple CNN structure, we are able to obtain $\sim 90\%$ accuracy in classification. The confusion matrix and receiver operating characteristic curve of classification demonstrate the strength and viability of this hybrid approach.
	\end{abstract}
	
	\section{Introduction}
	
	Identifying the type of dynamics that is experimentally measured is critical to time series classification, forecasting and statistical inference. This is because different analyses are required for different types of dynamics. In fact, the very essence of a time series is different between a stochastic process, a periodic process or a chaotic process. There has been much work recently to identify different dynamical processes using recurrence plots \cite{marwan2007recurrence}.  Recurrence plots were first introduced to visualize the recurrences of dynamical systems \cite{eckmann1987recurrence}. If $\{\vec{x}_i\}$, for $i=1,\cdots,N$, is a trajectory of a dynamical system with $N$ points, then the recurrence plot is a binary matrix defined as
	\begin{equation}
		\boldsymbol{R}_{ij} = 
		\begin{cases}
			1\text{, if $|\vec{x}_i-\vec{x}_j| \leq \epsilon$ }\\
			0\text{, if $|\vec{x}_i-\vec{x}_j| > \epsilon$}
		\end{cases},
	\end{equation}
	where $\epsilon$ is a threshold distance \cite{eckmann1987recurrence,marwan2007recurrence}. This concept has been widely applied in various fields such as astrophysics \cite{kurths1994testing,asghari2004stability,zolotova2006phase}, damage detection in engineering \cite{nichols2006damage}, molecular dynamics \cite{giuliani1996hidden,manetti2001recurrence}, economics \cite{orlando2017rqa,orlando2018recurrence,orlando2020business}, and medicine \cite{zbilut1990use,thomasson2001recurrence}.
	
	While recurrence plots and the associated statistical tools, such as spectral analysis \cite{stoica2005spectral} and Lyapunov exponents \cite{kantz1994robust,gao2006distinguishing,gao2013multiscale}, provide assistance in determining the nature of an unknown time series, they cannot be used to conclusively classify the dynamical system which generated the time series. In addition to being purely observational, many of the statistical tools require expensive amounts of computation or a large dataset in order to achieve high accuracy. 
	
	On the other hand, machine learning is a well established method for classification and forecasting, especially convolution neural networks (CNNs) for analysing images and matrices of measurements. Among the founders of the CNN we recall Hubel and Wiesel who proposed a cascading neural model for pattern recognition \cite{hubel1959receptive}, Fukushima that introduced  "neocognitron" as organized in two layers, the first for convolution and the second for downsampling  \cite{fukushima1982neocognitron}, and Atlas et al. who proposed a "dynamic formal neuron" (i.e. a temporal generalization of the formal neuron) for learning dynamic patterns \cite{atlas1987artificial}. 
	
	To fill the gap in classifying time series to different dynamical systems, we propose using a CNN trained on the recurrence plots. Recently, many applications combining recurrence plots with CNNs have been presented, such as recognition of physical activity \cite{garcia2018classification}, detection of Parkinson's disease \cite{afonso2019recurrence}, forecasting residential energy loads to optimize renewable energy resources \cite{estebsari2020single} and emotion recognition from electroencephalograms \cite{yang2018recurrence}. While those applications combining recurrence plots and CNNs have provided more insights on the matter, there has been no study on how to use time series to classify the nature of dynamical systems. In this paper, we tackle the problem by presenting a hybrid method for classifying different types of dynamics, such as stochastic, periodic and chaotic, using both recurrent matrices and CNNs.
	
	The main advantages of this hybrid approach are scalability, practicality and simplicity. Recurrence plots are not computationally expensive to generate, particularly for shorter time series, and CNNs are also easy and inexpensive to operate once trained. Even for training, there exist multiple software implementations in various programming languages that assist in quick prototyping. Furthermore, the ability to classify the underlying dynamical system using a small number of points is particularly useful for economic and historic data, which are limited to short time series. Finally, this hybrid method presents a simple and efficient way that is accessible to many non-specialist users, albeit there is a risk of misuse.

	\section{Literature review} \label{Sec:Literature}
	
	Time series classification (TSC) and, more generally, the problem of identifying the nature of a dynamical system (e.g. stochastic or deterministic) has attracted much attention. On TSC, among the great quantity of methods \cite{fawaz2019deep}, one can mention the use of the nearest neighbor (NN) classifier coupled
	with a distance function such as the Dynamic Time Warping (DTW). This constitutes a very strong baseline and, when compared to several distance measures, it has been shown that there is no single distance measure that significantly outperforms it \cite{lines2015time}. 
	
	Given this, the research has redirected the focus on ensembling methods to outperform the NN classifier coupled with DTW (NN-DTW).
	The idea is to pre-process time series so that the original data are transformed into a new feature space. After the pre-processing phase, ensemble methods are applied.
	%
	%
	Those techniques fall within three classes: bag-of-patterns, shape-based, and structure-based.

	\emph{Bag-of-patterns (BOP)} techniques that extract substructures of a time series as higher-level features, transform them via a method called  Symbolic Aggregate approXimation (SAX) \cite{lin2007experiencing} and uses the Euclidean distance (ED) as a similarity metric. A variant is the so-called bag-of-SFA-Symbols (BOSS) which``combines the extraction of substructures with the tolerance to extraneous and erroneous data using a noise reducing representation of the time serie'' \cite{schafer2015boss}.
	
	\emph{Shape-based} techniques combine  similarity metric with 1-nearest-neighbour (1-NN) classification \cite{vlachos2002discovering, rakthanmanon2012searching} and is used as a benchmark \cite{ding2008querying}. However, ``the problem with shape-based techniques is that they fail to classify noisy or long data containing characteristic substructures'' \cite{schafer2015boss}.
	
	\emph{Structure-based} techniques either extract higher-level features or they build a model with classical data mining algorithms like support vector machines (SVMs), decision trees, or random forests \cite{schafer2012sfa, bostrom2015binary, kate2016using}. Shapelets also belong to this class (i.e. classifiers that extract representative variable-length sequences from a time series)  \cite{hills2014classification, abadi2016tensorflow} or DTW features \cite{kate2016using}. The characteristic of the latter techniques over the others of this class is that they do not transform the entire data set, but instead identify a suitable partition and outperform the NN coupled with DTW (NN-DTW) \cite{bagnall2017great}. Because of this promising result, research focussed on the development of an ensemble, the so-called Collective Of Transformation-based Ensembles (COTE) classifiers \cite{bagnall2015time} ``that does not only ensemble different classifiers over the same transformation, but instead ensembles different classifiers over different time series representations'' \cite{fawaz2019deep}. 
	A further development is the extended COTE with a Hierarchical Vote system (HIVE-COTE) that uses a hierarchical structure with probabilistic
	voting. The latter is currently considered the state-of-the-art algorithm with regard to time series classification \cite{bagnall2017great, fawaz2019deep}.
	
	As mentioned, HIVE-COTE represents the state of the art but has become very computationally intensive and impractical  \cite{bagnall2017great}. For example, one of the classifiers is the shapelet transform \cite{hills2014classification} whose time complexity is $O(n^2 l^4)$, where $n$ is the number of time series in the dataset and $l$ is the length of a time series. Moreover, ``adding to the huge runtime of HIVE-COTE, the decision taken by 37 classifiers cannot be interpreted easily by domain experts, since researchers already struggle with understanding the decisions taken by an individual classifier'' \cite{fawaz2019deep}.
	
	Given the impracticability of HIVE-COTE algorithm, research has been redirected elsewhere. For example, following the idea of feature extraction and subsequent classification of time series, recurrence patterns have been identified by the means of a recurrence plot (RP) and its quantification (RQA) \cite{webber2015recurrence}. This is because, ``evidence suggest that recurrences contain all relevant information about a system's behaviour'' \cite{marwan2007recurrence} and ``the explicit representation of such regularities can reveal the underlining mechanisms that generated the data, and
	thus it is a potentially useful feature to classify time series'' \cite{silva2013time}. Another advantage of RP/RQA is the ability to identify segments of similar trajectories at arbitrary positions in multivariate time series as well as ``the dynamical properties, such as determinism, which reflect the pairwise (dis)similarity'' \cite{spiegel2014recurrence}. This approach has been proved superior to the classical dynamic time warping distance \cite{spiegel2014recurrence}.
	More recently, the spatial bag-of-features (SBoF) model and the deep convolutional neural networks (CNN) have been used for forecasting \cite{li2020forecasting} and successfully compared with automated algorithms encompassing autoregressive integrated moving average  (ARIMA), exponential smoothing algorithm (ETS), feed-forward neural network with autoregressive inputs (NNET-AR), exponential smoothing state space model with a Box-Cox transformation (TBATS), seasonal and trend decomposition using LOESS with AR modeling of the seasonally adjusted series (STLM-AR), random walk with drift (RW-DRIFT), theta method (THETA), naïve (NAIVE), and seasonal naïve (SNAIVE)  \cite{li2020forecasting, hyndman2015forecasting}. 
	
	Along these lines, we focus on the nature of a dynamical system trying to understand whether by means of feature extraction (RP) and classifiers (CNN) if it is random, chaotic or periodic. The potential of the suggested approach was recently demonstrated by a comparative analysis of noisy time series \cite{kirichenko2019time}. 
	
	\section{Data and material} \label{Sec:DataMat}
	
	\subsection{Time series data and generation of recurrence plots} \label{Sec:Data}
\begin{figure} [ht]
		\centering
		\includegraphics[width=0.5\linewidth]{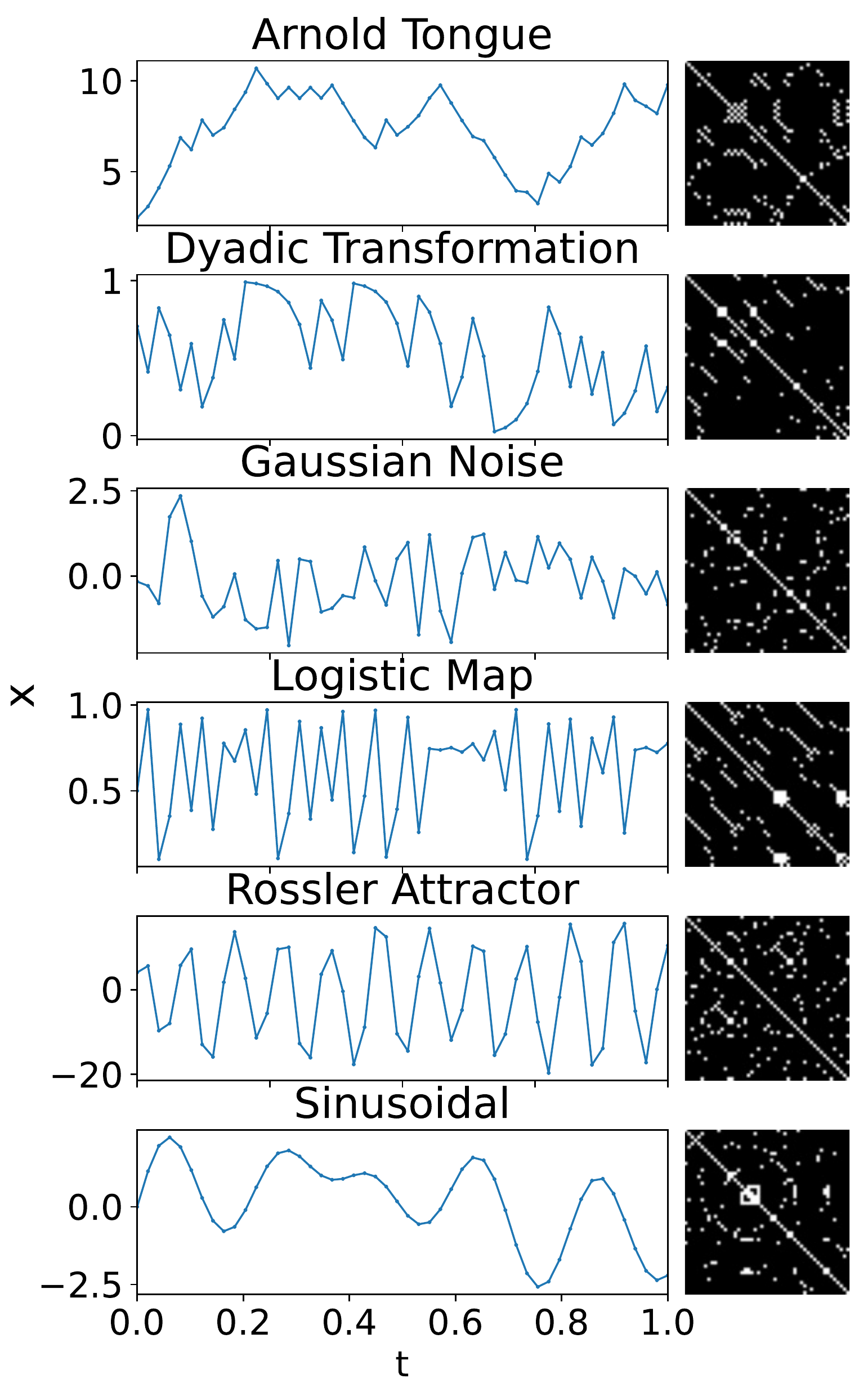}
		\caption{The trajectories (left) of simulated time-series containing 50 points and the recurrent matrices (right) generated from those trajectories.}
		\label{fig:Trajectories}
	\end{figure}

	Our data consists of 1,000 time series generated for each system under consideration (i.e. purely stochastic, periodic and chaotic). The systems considered are the Arnold tongue, dyadic transformation, Gaussian noise, logistic map, Rossler attractor and multi-frequency sine waves. These time series, of 50 data points each, are then processed with the toolbox CRP (R32.6) \cite{CRP522}. This is to obtain a recurrence plot (see Figure \ref{fig:Trajectories}).
	
	\subsection{Convolution neural network} \label{Sec:CNN} 
	
\begin{figure}
		\centering
		\includegraphics[width=0.5\linewidth]{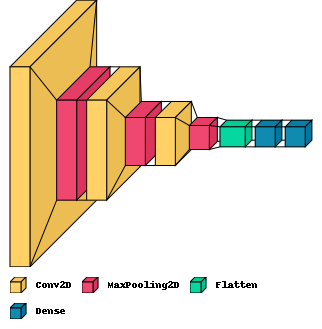}
		\caption{A diagram showing the CNN structure consisting of three convolution layers with 12 filters using a 5 by 5 kernel followed by three max-pooling layers of stride 2 and finally a densely connected network section to produce categorical predictions of different types of trajectories. This figure was generated using the `visualkeras' package.}
		\label{fig:CNN}
	\end{figure}

	To classify a given recurrence plot from time series data, we designed a CNN composed of three convolution layers with 12 filters using a 5 by 5 kernel each followed by max-pooling layers of stride 2 and finally a densely connected network section (see Figure \ref{fig:CNN}). The CNN was trained and tested using Tensorflow in Python3.
	
	\section{Results} \label{Sec:Results}
	
\begin{figure}
		\centering
		\includegraphics[width=1\linewidth]{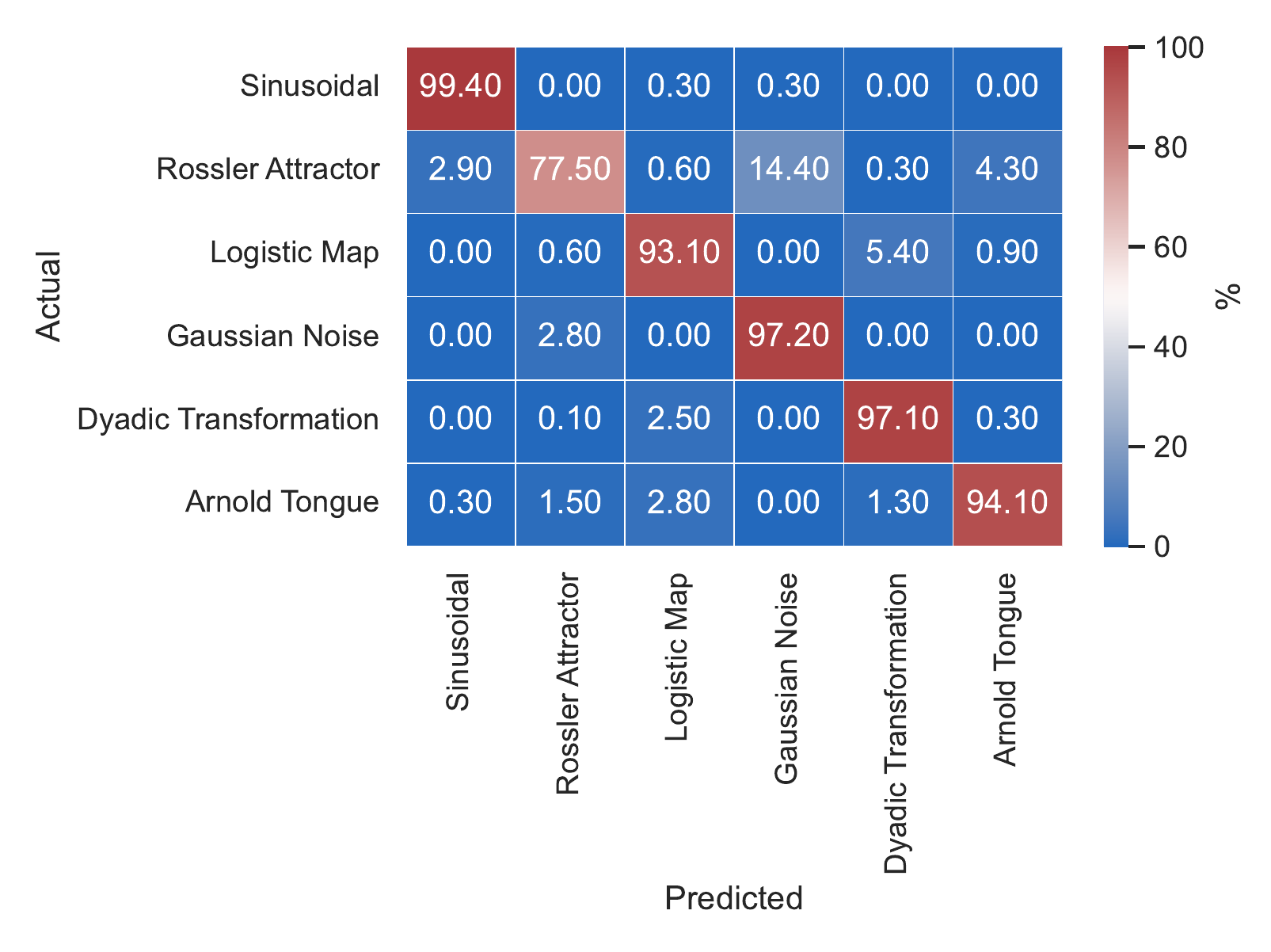}
		\caption{The confusion matrix of the CNN with 12 filters and 5 by 5 convolution kernel size. All but one dynamics achieve a success score above 90\%}
		\label{fig:ConfusionMat}
	\end{figure}

	Figure \ref{fig:ConfusionMat} shows the confusion or error matrix \cite{stehman1997selecting} that allows visualization of the performance of the CNN. The columns represent the instances in a predicted class and the rows represent the instances in an actual class. As shown, over six types of time series the prediction is 77.5\% correct at worst and 99.4\% correct at best. 
	
\begin{figure}
		\centering
		\includegraphics[width=1\linewidth]{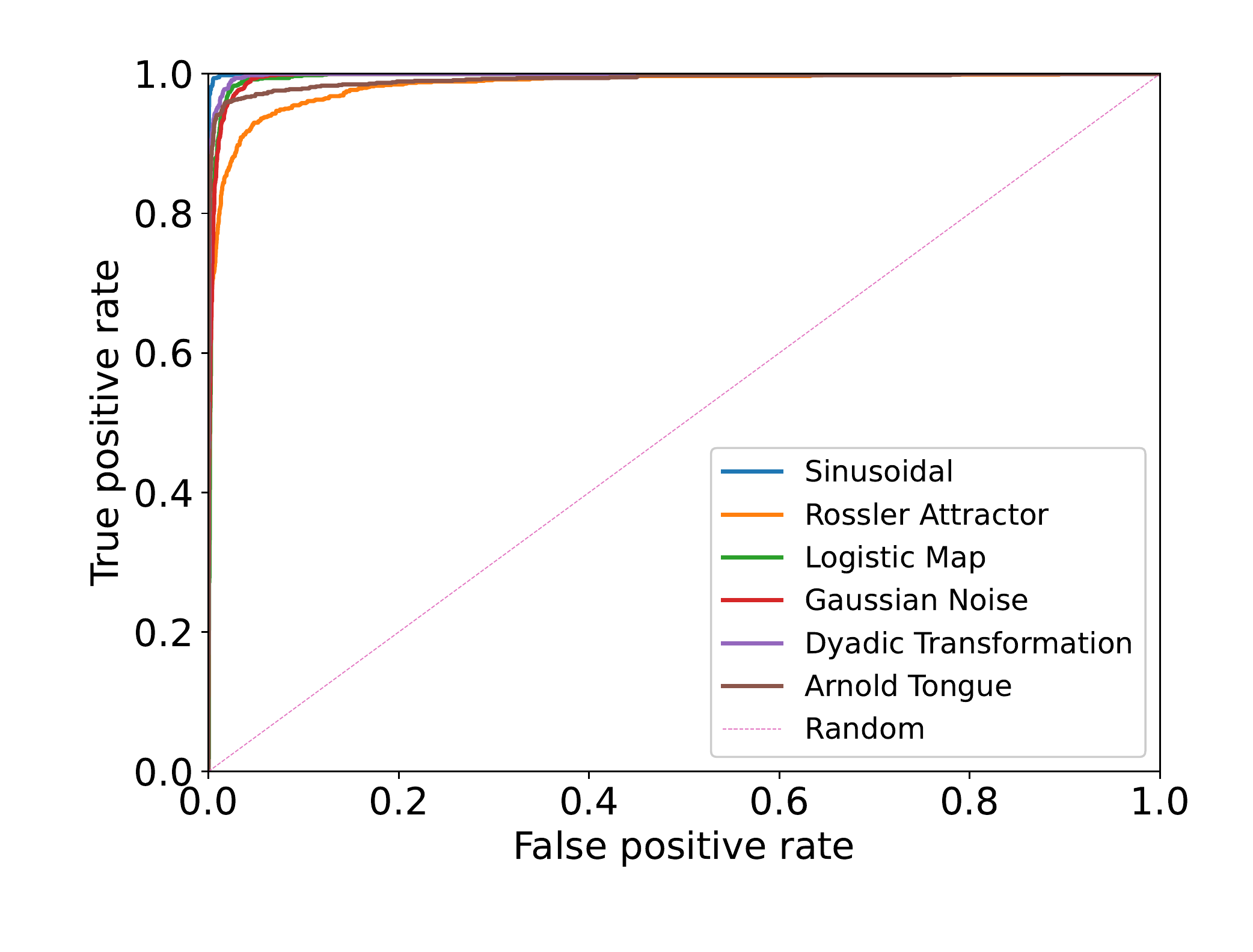}
		\caption{Multi-class ROC for each type of dynamic system. The true positive rate increases very rapidly for low levels of false positive rate, thus confirming the effectiveness of the suggested approach}
		\label{fig:ROC}
	\end{figure}

	A receiver operating characteristic curve (ROC) expresses the diagnostic ability of a binary classifier system \cite{fawcett2006introduction}. The ROC curve displays the true positive rate (TPR) versus the false positive rate (FPR) by changing the threshold. Figure \ref{fig:ROC} shows that the TPR is quite high for low levels of thresholds thus confirming the capability of the suggested approach in classifying recurrence plots of time series. The `Area under the ROC Curve' (AUC) is $0.994$ for `one-vs-one'  multiclass classification \cite{hand2001simple}.
	
	\section{Conclusion}
	In this work, we have demonstrated the effectiveness of using both RPs and CNNs in classifying the nature of dynamical systems from the generated time series. Our results show that this hybrid approach achieves very high accuracy despite only inputting 50 data points for a given time series and as obtained by the related RP. The rather high accuracy of the proposed classification is in agreement with the recent literature aimed at comparing noisy time series \cite{kirichenko2019time}. Furthermore, the RPs and CNN is fully scalable in terms of both parallel computing and GPU computing. Moreover, extraction of features through RP provides a visual and understandable classification by domain experts.
	
	The application of this research is in some domains such as economics and particularly in business cycles. In fact, they are nonlinear and irregular \cite{Brock1988}, However, most of the research concluded that chaotic models are sometimes theoretically fascinating but hold little to no practical use \cite{orlando2020business}. Notwithstanding, some found that a chaotic model could fit well with true data \cite{Orlando2016, Orlando2018a, Orlando2019Jun} including crashes such as those caused by the COVID-19 pandemic \cite{Orlando2019Jun}. 
	
	Thus, this research could provide the link between economic theory and identification of real dynamics based on machine learning classifiers and RQA/RP feature extraction where the latter has been successfully used to discover hidden dynamics and structural changes in economics \cite{orlando2017rqa, orlando2018recurrence}. 
	
	
	\bibliographystyle{unsrt} 
	
	\bibliography{main}
	
\end{document}